\newif\iflipics
\lipicsfalse

\newif\iflncs
\lncsfalse

\newif\ifsubmission
\submissionfalse

\iflipics
\documentclass[letterpaper,USenglish]{lipics-v2018}
\usepackage{graphics, subfiles, caption, wrapfig, xspace}
\usepackage{microtype}
\nolinenumbers
\else\iflncs
\documentclass{llncs}

\usepackage{graphics, subfiles, wrapfig, xspace,amsmath, amsfonts, amsthm,amssymb, color}
\else
\documentclass[11pt]{article}
\usepackage{amsmath, amsfonts, amsthm, amssymb, color,graphics, subfiles, caption, wrapfig, xspace}
\usepackage[plain]{fancyref}
\allowdisplaybreaks
\usepackage{fullpage}
\usepackage{hyperref}
\newtheorem{lemma}{Lemma}
\newtheorem{theorem}[lemma]{Theorem}
\newtheorem{definition}{Definition}

\fi\fi

\allowdisplaybreaks

\newcommand{\enm}[1]{\ensuremath{#1}\xspace}
\newcommand{\abs}[1]{\left\lvert #1 \right\rvert}
\newcommand{\opt}{\enm{\textsf{OPT}}}

\newcommand{\alg}{\enm{\textsf{ALG}}}
\newcommand{\ucp}{\enm{\textsc{UCP}}}
\newcommand{\pds}{\enm{\textsc{MaxPDS}}}
\newcommand{\scrS}{\enm{\mathcal{S}}}
\newcommand{\bbE}{\enm{\mathbb{E}}}
\newcommand{\bbZ}{\enm{\mathbb{Z}}}
\newcommand{\p}[1]{\enm{\text{Pr}\left[ #1 \right]}}

\newcommand{\e}[1]{\enm{\bbE\left[ #1 \right]}}
\newcommand{\ev}[2]{\enm{\bbE_{#1}\left[ #2 \right]}}
\newcommand{\bbN}{\enm{\mathbb{N}}}
\newcommand{\set}[1]{\enm{\left\lbrace #1 \right\rbrace}}

\newcounter{newnote}[section]
\iflncs
\renewcommand{\thenote}{\thesection.\arabic{newnote}}
\fi

\newcommand{\mdnote}[1]{\refstepcounter{newnote}$\ll${\color{red} \bf Mike's Comment~\thenote:}
{\sf \color{red} #1}$\gg$\marginpar{\tiny\bf MD~\thenote}}
\newcommand{\nnote}[1]{\refstepcounter{newnote}$\ll${\color{blue} \bf Naomi's Comment~\thenote:}
{\sf \color{blue} #1}$\gg$\marginpar{\tiny\bf NE~\thenote}}

\iflipics
\renewcommand{\mdnote}[1]{}
\renewcommand{\nnote}[1]{}
\fi
\iflncs
\renewcommand{\mdnote}[1]{}
\renewcommand{\nnote}[1]{}
\fi

\iflipics
\title{Reception Capacity: Definitions, Game Theory and Hardness}
\author{Michael Dinitz}{Johns Hopkins University, Baltimore, MD, USA}{mdinitz@cs.jhu.edu}{}{}
\author{Naomi Ephraim}{Cornell University, Ithaca, NY, USA}{nephraim@cs.cornell.edu}{}{}
\authorrunning{M. Dinitz and N. Ephraim}
\Copyright{Michael Dinitz and Naomi Ephraim}
\subjclass{Theory of Computing $\rightarrow$ Quality of Equilibria, Mathematics of Computing $\rightarrow$ Approximation Algorithms}
\else
\iflncs
\title{Reception Capacity: Definitions, Game Theory and Hardness}
\author{Michael Dinitz\inst{1}\thanks{Supported in part by NSF awards CCF-1464239 and CCF-1535887} \and Naomi Ephraim\inst{2}\thanks{Supported in part by NSF Award SATC-1704788 and NSF Award IIS-1703846}}
\institute{Johns Hopkins University, Baltimore MD, USA \\ \texttt{mdinitz@cs.jhu.edu} \and Cornell University, Ithaca NY, USA \\ \texttt{nephraim@cs.cornell.edu}}
\else
\title{Reception Capacity: Definitions, Game Theory and Hardness}
\author{Michael Dinitz\thanks{Supported in part by NSF awards CCF-1464239 and CCF-1535887} \\ Johns Hopkins University \and Naomi Ephraim\thanks{Supported in part by NSF Award SATC-1704788 and NSF Award IIS-1703846} \\ Cornell University}
\date{}
\fi
\fi
\begin{document}

\let\temp\newpage
\let\newpage\relax
\maketitle

\begin{abstract}
The \emph{capacity} of wireless networks is a classic and important topic of study.  Informally, the capacity of a network is simply the total amount of information which it can transfer.  In the context of models of wireless radio networks, this has usually meant the total number of point-to-point messages which can be sent or received in one time step.  This definition has seen intensive study in recent years, particularly with respect to more accurate models of radio networks such as the SINR model.  This paper is motivated by an obvious fact: radio antennae are (at least traditionally) omnidirectional, and hence point-to-point connections are not necessarily the best definition of the true \emph{capacity} of a wireless network.  To fix this, we introduce a new definition of \emph{reception capacity} as the maximum number of messages which can be received in one round, and show that this is related to a new optimization problem we call the Maximum Perfect Dominated Set (MaxPDS) problem.  Using this relationship we give a tight lower bound for approximating this capacity which essentially matches a known upper bound. As our main result, we analyze this notion of capacity under game-theoretic constraints, giving tight bounds on the average quality achieved at any coarse correlated equilibrium (and thus at any Nash).  This immediately gives bounds on the average behavior of the natural distributed algorithm in which every transmitter uses online learning algorithms to learn whether to transmit.
\end{abstract}

\iflipics
\newpage 
\else
\thispagestyle{empty}
\newpage 
\fi

\let\newpage\temp

\section{Introduction}
A fundamental quantity of a wireless network is its \emph{capacity}, which informally is just the maximum amount of data which it can transfer.  There is a large literature on analyzing and computing the capacity of wireless networks under various modeling assumptions, including models of how interference works and assumptions on how nodes are distributed in space.  The last decade has witnessed a flurry of activity in this area, particularly for worst-case (rather than random) node distributions, motivated by the ability to apply ideas from multiple areas of theoretical computer science (approximation algorithms and algorithmic game theory in particular) to these problems.  

We continue that line of work in this paper, but with a new definition of capacity.  Much of the research in the last decade (see, e.g.,~\cite{GHW14,GWHW09,dinitz09,HM14,HM12,HM11,Kes11,HT15}) has used a point-to-point definition of capacity: given a collection of disjoint pairs $(s_i, t_i)$ of nodes (called the demands), and some model of interference, the point-to-point capacity is the maximum number of pairs which can simultaneously successfully transmit a message from $s_{i}$ to $t_{i}$.  This is sometimes motivated by its utility in scheduling: if we are trying to support many unicast demands in a wireless network, a natural thing to do is make as much progress as possible in each time step, i.e., maximize the number of successful transmissions.  

But while well-motivated by scheduling, this is not the only possible definition of \emph{capacity}.  In particular, a natural notion of ``capacity" is of the best case: what is the absolute limit on the usefulness of a network in even the best possible situation?  With this intuition, there are two main issues with point-to-point capacity: the existence of demands, and the requirement for unicast communication.  First, since we want to talk about the capacity of a network, why should the capacity be a function of any set of input demands (which are, after all, external to the network itself)?  Instead we should allow any set of demands and take the best possible.  
So one might instead define the ``capacity'' of the network  to be the maximum number of $(s_i, t_i)$ pairs which can simultaneously successfully transmit a message, but not restrict $(s_i, t_i)$ to come from any particular input subset (or equivalently, require the set of input demands to always be $V \times V$ where $V$ is the set of nodes).

Even if we remove the demands, though, there is still something restrictive about this notion of capacity: it only allows unicast, point-to-point communication.  One of the defining features of traditional wireless networks is that antennas are omnidirectional. 
Thus, if we want to truly understand the ``capacity'' of a given wireless network, we should surely take into account the ability for a single node to successfully send the same message to many other nodes in one time slot, since in the best case we can obtain significant benefits from this ability.  

For example, suppose we are in a classical radio network represented by a communication graph, where each node is a transmitter who can communicate with its neighboring nodes. In this model, interference is destructive: $u$ will receive a message from $v$ if $v$ sends a message, $u$ does not send a message, and no other neighbor of $u$ sends a message. 
Suppose that we are given a star topology with $r$ as the center and leaves $x_1, \dots, x_{n}$.  What is the capacity of this network?  Traditionally, the answer would be $1$: only one of the unicast links can be successful, since $r$ can only send or receive one message at a time.  On the other hand, if $r$ really only has a single message which it is trying to send to all of its neighbors, then there can be $n$ successful receptions in a single round, and hence the capacity should be $n$.  


Motivated by this, we define a new notion of capacity in radio networks which we call the \emph{reception capacity}.  
Informally, this is simply the maximum number of successful message receptions in a single round.  Note that there are no demands, and there is no requirement that different receptions correspond to different messages. Hence this definition is the true limit on the single-step ``usefulness'' of the network.  We emphasize that there are many notions of capacity, each of which is appropriate and interesting in different contexts, and we are not claiming that reception capacity is the \emph{right} definition.  We are merely claiming that it is a natural definition of ``best-case usefulness", so bounds on it are bounds on the utility of a network even in the best possible situation.

In this paper we study this notion of capacity in radio networks. We first show that maximizing this capacity is equivalent to a new optimization problem we call the \emph{Maximum Perfect Dominated Set} (\pds) problem.  While this problem as defined is new, we show that the classical Decay protocol of~\cite{BGI92} gives an $O(\log n)$-approximation algorithm. We also give a tight lower bound on its approximability which matches this upper bound, under plausible complexity assumptions. Both of these results are with respect to networks defined by general communication graphs (the networks are not restricted to having any specific structure). Together, these two bounds give us a tight understanding of the approximability of maximizing the reception capacity. 

The main technical contribution of this paper, though, is the study of the capacity achieved by \emph{self-interested} agents.  What if every transmitter has its own goals, which do not necessarily align with the global objective of maximizing the reception capacity?  While there are many ways to model this, we take a first step by considering a natural model in which every transmitter wants to broadcast its message to as many of its neighbors as possible, but is penalized for unsuccessful transmissions.  This intuitively corresponds to a setting where transmitters want to get their message out to many of their neighbors (e.g., if it is an important piece of information which the transmitter wants to disseminate) but are discouraged from placing an unnecessary load on the network if there will be many unsuccessful transmissions. 

This type of setting is naturally modeled as a game, where each transmitter is a player that is trying to maximize its own utility.  In such a game, what can we say about the \emph{achieved} reception capacity?  Does the selfishness of the transmitters mean the network is being underutilized, or do they naturally arrive at an equilibrium with close to optimal reception capacity?  In the unicast setting, Dinitz~\cite{dinitz10} showed that equilibria of the related unicast-specific game can be arbitrarily far from optimal: can the same thing happen with reception capacity?

We completely characterize the behavior of a broad class of equilibria known as \emph{$\epsilon$-coarse correlated equilibria} ($\epsilon$-CCE), which both generalize Nash equilibria and to which natural distributed online learning algorithms (known as no-regret algorithms) will converge~\cite{BHLR08}.  In particular, for a network with $n$ nodes, we prove that at \emph{every} $\epsilon$-CCE the achieved reception capacity of the transmitters is at least $\Omega(1/\sqrt{n})$ of the true reception capacity (unlike the unicast setting), and moreover there exist instances in which \emph{every} $\epsilon$-CCE achieves reception capacity that is at most $O(1/\sqrt{n})$ of the true reception capacity.  


\subsection{Modeling}

To model the notion of reception capacity, we consider the classical \emph{radio network} model.  In this model there is a communication graph $G= (V, E)$, and each node in $V$ can act as either a transmitter or a receiver.  In a given unit of time (we make the standard assumption of synchronous rounds), each node can either broadcast a message to all of its neighbors, or choose to not broadcast and thus act as a receiver.  Interference is modeled by requiring that a receiver can only receive one message in each round, or else the messages interfere and cannot be decoded.  In other words, a vertex $i$ can successfully decode a message from a neighbor $j$ if and only if $i$ is not broadcasting (and so is acting as a receiver), $j$ is broadcasting, and no other neighbor of $i$ is broadcasting.  If multiple neighbors of $i$ are broadcasting then their messages all interfere with each other at $i$, and so $i$ would not receive any message.  

In this model, the equivalent of the unicast notion of ``capacity'' used in recent work would be a maximum induced matching (or if there is a set of input demands, a maximum induced matching subject to being a subset of the demands).  This is because, in the unicast setting, each node can only transmit to a single neighbor or receive a message from a single broadcasting neighbor. Therefore, maximizing the unicast capacity is equivalent to finding a set $S$ of broadcasters and a set $T$ of receivers such that the bipartite subgraph induced by $S$ and $T$ is a matching, and maximizing the size of this matching. 

However, this may be significantly smaller than the number of nodes which can successfully hear a message, as the star example shows.  So we will instead adopt a different notion of capacity:

\begin{definition} \label{def:capacity}
The \emph{reception capacity} of a wireless network 
is the maximum number of nodes which can simultaneously successfully receive a message.
\end{definition}

We note that this differs from the traditional unicast or multicast setting because there are no demands from broadcasters to receivers. The reception capacity is rather the total number of messages that can be received in one round, without any assumptions on whether one node ``wants'' to send a message to another node. Thus it is a true upper bound on the ``capacity" (usefulness) of the network.  


\section{Our Results} 

\subsection{MaxPDS and Approximability}

We first observe that it is straightforward to relate reception capacity to reasonably well-studied notions in graph theory.  In particular, since each node successfully receives a message if and only if it does not broadcast and exactly one of its neighbors does broadcast, we would like each receiver to be \emph{perfectly dominated} by the set of broadcasting nodes. 

\begin{definition} \label{def:perfectly-dominated}
Given a graph $G= (V, E)$ and a set $S \subseteq V$, we say that a node $v \in V \setminus S$ is \emph{perfectly dominated} by $S$ if there exists exactly one node $u \in S$ such that $u$ is a neighbor of $v$. 
\end{definition}
For every subset $S \subseteq V$, let $D(S) = \{v : v \text{ is perfectly dominated by } S\}$.  This immediately lets us relate the reception capacity to perfect domination.
\begin{lemma}
The reception capacity of a wireless network $G = (V, E)$ is exactly $\displaystyle \max_{S \subseteq V} |D(S)|$.
\end{lemma}
\begin{proof}
Let $S \subseteq V$.  If every node in $S$ broadcasts a message, by the definition of the radio network model, a node receives a message if and only if it is in $D(S)$.  Hence the reception capacity is at least $\max_{S \subseteq V} |D(S)|$.  On the other hand, let $S$ be the set of nodes who transmit when the reception capacity is achieved, and suppose that $v$ receives a message.  Then $v \in D(S)$, and hence the reception capacity is at most $\max_{S \subseteq V} |D(S)|$.  
\end{proof}
Thus computing the reception capacity of a network is equivalent to the following optimization problem.
\begin{definition}
Given a graph $G = (V, E)$, the \textsc{Maximum Perfect Dominated Set Problem} (\pds) is to find a set $S \subseteq V$ which maximizes $|D(S)|$.  
\end{definition}
This problem seems to be new, despite the vast literature on variations of dominating sets.  It is superficially similar to the well-studied \textsc{Minimum Perfect Dominating Set} problem~\cite{W94,YL90}, in which the goal is to find the set $S$ of minimum size such that $D(S) = V \setminus S$ (note that some such $S$ always exists since we could set $S = V$).  Despite their superficial similarity, though, the problems are quite different: in \pds nodes not in $S$ may still not be perfectly dominated, so both the feasible solutions and the objective functions of the two problems are quite different.  


Therefore, our first goal is to characterize the hardness of \pds. We observe that the classical \emph{Decay protocol}~\cite{BGI92} can be used to obtain a simple $O(\log(n))$ approximation algorithm for \pds. 
We compliment this with an essentially matching lower bound for \pds.
The precise lower bound depends on the hardness assumption, but all are essentially polylogarithmic.  
\begin{theorem} \label{thm:hardness}
\pds cannot be approximated to better than a polylogarithmic factor.  More precisely:
\begin{itemize}
\item Let $\varepsilon > 0$ be an arbitrary small constant. Suppose that NP $\not \subseteq \text{BPTIME}(2^{n^{\varepsilon}})$.  Then there is no polynomial time algorithm which approximates \pds to within $O(\log^{\sigma} n)$ for some constant $\sigma = \sigma(\varepsilon)$.  
\item Under Feige's Random 3SAT Hypothesis~\cite{Feige02}, no polynomial time algorithms approximates \pds to within $O(\log^{1/3 - \sigma} n)$ for arbitrarily small constant $\sigma > 0$.  
\item Under the assumption that the Balanced Bipartite Independent Set Problem (BBIS) cannot be approximated better than $O(n^{\varepsilon})$ for some constant $\varepsilon > 0$ (Hypothesis 3.22 of~\cite{Demaine08}), there is no polynomial time algorithm which approximates \pds to within $o(\log n)$.  
\end{itemize}
\end{theorem}

This lower bound is obtained through a connection to the \emph{Unique Coverage Problem} (\ucp).  
Informally, \ucp is a variation of Maximum Coverage with a similar uniqueness requirement as in \pds (an element only counts as covered if it is contained in exactly one chosen set).  Upper and lower bounds for \ucp are known~\cite{Demaine08}, so we derive our lower bound by reducing from \ucp to \pds (in particular, the different lower bounds and their hardness assumptions are all direct from equivalent bounds and assumptions for \ucp). The lower bound is given in Section~\ref{sec:hardness} and the upper bound is given in Appendix~\ref{sec:approx}.

\subsection{Reception Capacity with Self-Interested Agents}

The above algorithmic results provide us with a comprehensive understanding of the problem of maximizing the reception capacity in arbitrary radio networks. However, they do not imply bounds on the usability of these networks with respect to their reception capacity. That is, for a given network, we would like to investigate the capacity that is utilized under reasonable behavioral assumptions. We focus on the model of self-interested agents due to the competitive nature of a network with fully destructive interference, and because it is a tractable and standard model in the literature on algorithmic game theory. 

Therefore, the main focus of this paper is a natural game-theoretic formalization which we call the \emph{reception capacity game}.  Informally, this is a game in which the nodes are self-interested players, and the utility of each node is $0$ if it does not transmit, and otherwise is a linear function of the number of neighbors who successfully heard its message and the number who did not (we define this game formally in Section~\ref{sec:game}).  In other words, each node gets some positive utility from successfully transmitting its message to a neighbor, but pays a price for an unsuccessful transmission. 

While this game may seem somewhat arbitrary, it is quite natural.  Clearly there has to be some penalty for unsuccessful transmissions, or else the only equilibrium is for all nodes to broadcast all the time.  This motivated the previous work on unicast capacity in which a similar game is analyzed~\cite{dinitz09,dinitz10,AM11}, and in fact our game is the obvious generalization of the earlier unicast capacity game. It also motivated previous work on clique networks~\cite{FYN07}, where they analyzed equilibria in which all nodes were required to transmit with probability strictly smaller than 1. Thus, while it may not be a perfect model of the incentives of selfish transmitters, it is reasonable in at least some situations (e.g., if every transmitter is trying to broadcast an advertisement of some kind). More importantly, it provides insight into the limits of the performance of radio networks in the presence of self-interested agents. 

When we analyze behavior in a game, the natural approach is to study the quality of the solution at some notion of equilibrium (this is the well-studied notion of \emph{inefficiency of equilibrium} in algorithmic game theory).  While the most popular notion of equilibrium to study is the famous Nash equilibria, we provide stronger results by studying \emph{coarse-correlated equilibria (CCE)}, or more precisely, approximate versions known as $\epsilon$-CCE. We define these formally in Section~\ref{sec:game}, but CCE are generalizations of Nash equilbria, and hence if we can prove that all CCE are close to optimal, or if we can prove that all CCE are far from optimal, then these bounds immediately hold for Nash equilibria as well.  Moreover, CCE are an important class of equilibria in a distributed context since (unlike Nash equilibria) natural distributed learning algorithms will have an empirical distribution of play which converges to a CCE, and thus CCEs can be computed efficiently even in distributed settings.   We note that these equilibria are precisely those analyzed and used in~\cite{dinitz10,AM11} to design distributed algorithms for unicast capacity.  However, it was shown in~\cite{dinitz10} that in \emph{arbitrary} graphs, no nontrivial bounds were possible: there are examples in which there is a solution with $\Omega(n)$ successful transmissions, while any CCE has an average of at most $O(1)$ successful transmissions. On the other hand, we will prove that even in arbitrary graphs, the expected number of receptions in any CCE is at most an $O(\sqrt{n})$ factor worse than \opt (the true reception capacity).  We will also show that this is tight by designing instances in which \emph{all} CCE are $\Omega(\sqrt{n})$ worse than \opt. More formally, we prove the following theorems.

\begin{theorem} \label{thm:pota}
In any instance of the reception capacity game, the expected number of successful receptions in any $\epsilon$-CCE is at least $\opt \cdot \Omega\left(\frac{1}{\sqrt{n}} - \epsilon)\right)$. 
\end{theorem}

\begin{theorem} \label{thm:pots}
There is an instance of the reception capacity game in which in every $\epsilon$-CCE, the expected number of successful receptions is at most $\opt \cdot O((1 + \epsilon) / \sqrt{n})$. 
\end{theorem}

Note that since every Nash equilibrium is a $0$-CCE, our bounds immediately imply bounds on the more classical Price of Anarchy / Stability, in which we compare the optimal solution to the worst / best Nash.  We prove Theorem~\ref{thm:pota} in Section~\ref{sec:pota} and Theorem~\ref{thm:pots} in Section~\ref{sec:pots}.  

Interestingly, like the unicast capacity game studied in~\cite{dinitz09,dinitz10,AM11} but unlike most algorithmic game theory settings, our notion of ``quality'' is \emph{not} the social welfare, i.e., it is not just the sum of the utilities of the players.  Our notion of quality is number of successful receptions, which can be dramatically different from the social welfare.  This means that standard techniques such as \emph{smoothness}~\cite{roughgarden15} cannot be used to analyze this game.

\subsection{Related Work}

\subsubsection{Capacity in wireless networks} As discussed earlier, this paper follows an extensive line of work on computing the capacity of wireless networks.  There has been a particular focus on the SINR or physical model, in which we explicitly reason about the signal strength and interference at each receiver.  However, there has also been significant work directly on graph-based models (e.g., \cite{dinitz10}) and on the relationship between graph models and the SINR model~\cite{HT15} (which shows in particular that graphs can do a surprisingly good job of representing the physical model, motivating continued study of graph models).  

From the perspective of computing the capacity, the most directly related work (and much of the inspiration for this paper) are~\cite{dinitz09} and \cite{GWHW09}, which to a large extent introduced the unicast capacity problem for worst-case inputs and gave the first approximation bounds.  These bounds were improved in a series of papers, most notably including a constant-factor approximation~\cite{Kes11}, and have been generalized to even more general models and metrics, e.g.~\cite{HM11,HM14}.  

Much of this paper focuses on analyzing a natural game-theoretic version of reception capacity.  This is directly inspired by a line of work on a related game for unicast capacity, initiated by~\cite{dinitz09} and continued in~\cite{dinitz10,AM11}.  These papers study various equilibria for the unicast capacity game (Nash equilibria in~\cite{dinitz09}, coarse correlated equilibria in~\cite{dinitz10,AM11}) and prove what are essentially price of anarchy or total anarchy bounds (upper bounds on the gap between the optimal capacity and the capacity at equilibrium).  


\subsubsection{Radio networks} There is a long line of research on the radio network model under various assumptions. Much of this work focuses on the \emph{radio broadcast problem}~\cite{CK85,C91}, in which there is a graph representing the network and a source node $s$. The problem is to minimize the number of rounds that it takes for a message, originating at $s$, to be successfully propogated through the network. 

The literature on this model includes many algorithmic results. NP-hardness results were shown in~\cite{CK85,EGMT84}, approximation algorithms were given by~\cite{BGI92,C91,GM03,EK05a,GPX05}, and inaproximability results were given in~\cite{KM98,ABLP91,EK04,EK05b}.  
Despite the vast work on this problem, reception capacity differs from the radio broadcast problem in that there we are trying to determine the optimal set of broadcasters in each round, rather than determining a broadcasting schedule given a set of nodes who are allowed to transmit.

Nevertheless, some results in the radio network models apply to the case of reception capacity. In particular, the Decay protocol, introduced as a means of giving an approximation algorithm to the radio broadcast problem, yields an approximation algorithm for \pds~\cite{BGI92} (given in Appendix~\ref{sec:approx}). Another line of work which focuses on testing communication lines between nodes in networks provides results which imply that \pds is NP-hard. In particular, Even et al.\ show a reduction from a similar setting to a variant of the Exact Cover problem (which is a variant of Set Cover in which each element must be covered by exactly one set)~\cite{EGMT84}. Their proof can be used to show that the \pds on bipartite graphs is NP-hard, thus implying hardness for the general case. 

A notable variant of the radio broadcast problem is that of \emph{gossiping} in radio networks, which is sometimes called all-to-all communication~\cite{GPX05}. This problem studies the number of rounds for $n$ messages, one originating at each of $n$ nodes, to be propagated through the network. While this problem shares a closer resemblance to \pds than that of radio broadcasting, we are not aware of any results that directly imply results for \pds. 

\subsection{Notation}
Given any graph $G = (V,E)$, we refer to undirected graphs with $\abs{V} = n$. Additionally, for any vertex $v \in V$, we define $N(v)$ as the open neighborhood of $v$, that is, $N(v) = \{u \in V : \{v,u\} \in E\}$, and we let $d(v)$ denote the degree of $v$.

\section{Hardness of Approximation} \label{sec:hardness}
In this section, we present our hardness of approximation result  for the \textsc{Maximum Perfect Dominated Set Problem}. 
We begin by defining the Unique Coverage Problem~\cite{Demaine08}.

\begin{definition}[\cite{Demaine08}] Given a universe $U$ of elements and a collection \scrS of subsets of $U$, the \textsc{Unique Coverage Problem (UCP)} is to find a subcollection $S \subseteq \scrS$ of subsets which maximizes the number of elements that are uniquely covered, i.e., are in exactly one set of $S$.
\end{definition}


Demaine et al.~\cite{Demaine08} proved the equivalent of Theorem~\ref{thm:hardness} for \ucp (all bounds and assumptions are exactly the same, just for \ucp rather than \pds) and an $O(\log n)$-approximation for \ucp. Because of the similarity between \ucp and \pds, we base our lower approximability bound on \ucp, and in particular, show an approximation-preserving reduction from \ucp to \pds. 


\begin{theorem}
\label{thm:ha}
Assuming \ucp cannot be approximated to within $O(\log^{c}(n))$ for some constant $c$ satisfying Theorem~\ref{thm:hardness}, then \pds is hard to approximate to within $O(\log^{c}(n)).$ 
\end{theorem}

\begin{proof}
Consider an instance of UCP with a universe $U$ of elements and a collection $\scrS$ of subsets of $U$. For specified parameters $\alpha', \beta'$, given a subcollection $\scrS' \subset \scrS,$ we define the following two cases. 
\begin{enumerate}
\item $\scrS'$ is a Yes-instance of UCP if the number of elements uniquely covered is at least $\alpha'.$ 
\item $\scrS'$ is a No-instance of UCP if the number of elements uniquely covered is less than $\beta'.$
\end{enumerate}

Given an instance of this problem, construct an undirected bipartite graph $G' = (V',E')$ such that $V'$ consists of a vertex $s_{i}$ for each set $S_{i} \in \scrS$ and a vertex $x_{i}$ for each element $e_{i} \in U.$ Let $\{s_{i}, x_{j}\} \in E'$ if $e_{j} \in S_{i}.$ Let $A$ denote the set of vertices $s_{i}$ corresponding to sets in $\scrS,$ and let $B$ denote the vertices corresponding the elements in $U.$ 

Construct a new bipartite graph $G = (V,E)$ such that $V$ consists of $A$ and $k$ copies of $B,$ denoted $B_{1}, B_{2}, \ldots, B_{k}.$ Let $V$ have an additional vertex $v$ that is adjacent to all vertices in $A.$ Let $E$ consist of $k$ copies of $E',$ one for each bipartite subgraph over $(A, B_{i})$ for all $i \in [k]$.


Consider some solution $\scrS'$ to the UCP instance. Define $D = \{s_{i} : S_{i} \in \scrS'\} \cup \{v\}.$ If $\scrS'$ is a Yes-instance of UCP, then the number of vertices perfectly dominated by $D$ is $\alpha \geq \alpha' k,$ because in each of the $B_{i},$ there are at least $\alpha'$ perfectly dominated vertices. On the other hand, if $\scrS'$ is a No-instance of UCP, then there are only $\beta < \abs{\scrS} + k\beta'$ vertices perfectly dominated by $D,$ because $\{s_{i} : S_{i} \in \scrS'\}$ perfectly dominates less than $k\beta'$ of the vertices in the $B_{i}$ and $v$ perfectly dominates the $\abs{\scrS}$ vertices in $A.$ 

Now, set $k = \abs{\scrS}.$ Then $\alpha \geq \alpha' \abs{\scrS} = \alpha' k$ and $\beta < \abs{\scrS} + \abs{\scrS}\beta' = k + k\beta' = k(\beta' + 1)$.  Therefore, the approximation ratio for \pds in this setting is 
$\frac{\alpha}{\beta} > \frac{\alpha'k}{k(\beta' + 1)} = \frac{\alpha'}{\beta'+1} \geq \frac{\alpha'}{2\beta'}$ when $\beta' \geq 1$, which is trivially true. Since all we have done is create $\abs{\scrS}$ repetitions of $B$, this can be done in polynomial time. 

Therefore, this reduction begins with an instance of UCP with an approximation ratio of $\frac{\alpha'}{\beta'}$ and transforms the problem into an instance of \pds with an approximation ratio of $\frac{\alpha}{\beta}.$ Let $n'$ be the size of the input to this reduction, and let $n$ be the size of the resulting instance of \pds. By assumption, $\frac{\alpha'}{\beta'} = \Omega(\log^{c}(n')).$ Therefore, we want to show that $\frac{\alpha}{\beta} = \Omega(\log^{c}(n)).$ We start with $n' = \abs{\scrS} + \abs{E'}$ and we end with $n = \abs{\scrS} + k\abs{E'}$.  Then 
$n = \abs{\scrS} + k\abs{E'} = \abs{\scrS} + \abs{\scrS} \abs{E'} = \abs{\scrS}(1 + \abs{E'}) \leq 2\abs{\scrS} \abs{E'} < 2(n')^{2}$, and hence 
$\log^{c}(n) \leq \log^{c}(2(n')^{2}) \leq 4^{c}\log^{c}(n')$, implying that 
$\log^{c}(n') \geq \frac{1}{4^{c}}\log^{c}(n).$ Therefore, 
$\frac{\alpha}{\beta} \geq \frac{\alpha'}{2\beta'} \geq \frac{1}{2}\log^{c}(n') \geq \frac{1}{4^{c+1}}\log^{c}(n) = \Omega(\log^{c}(n))$ as desired, thus showing that \pds is hard to approximate to within $O(\log^{c}(n))$. 
\end{proof}

This reduction from \ucp to \pds shows that \pds is hard to approximate to within $O(\log^{c}(n))$ under any hardness assumption for which \ucp is hard to approximate to within $O(\log^{c}(n))$. In particular, this holds for the three different hardness assumptions used to show the hardness of approximating \ucp in~\cite{Demaine08}, thus proving Theorem~\ref{thm:hardness}.

\section{The Reception Capacity Game}
\label{sec:game}
In this section, we study reception capacity as a game in a distributed setting with self-interested players. 
Formally, an instance of the \textit{reception capacity game} is represented by a graph $G = (V,E)$, where we let $V = [n]$ represent the players. Each player has two actions: broadcast (represented by $1$) or be silent (represented by $0$).   Let $S = \{0,1\}^{n}$ be the strategy space, where for each $s \in S$, $s_{i}$ represents the action of player $i$ for each $i \in [n]$. For any $s$, if $s_{i} = 1$ define
$r_{i}(s) = \abs{\set{j \in N(i) : s_{j} = 0 \wedge \sum_{k \in N(j)} s_{k} = 1}}$ as the number of neighbors of $i$ not broadcasting and receiving exactly one message under 
$s$, and if $s_{i} = 0$, let $r_{i}(s) = 0$. That is, when $i$ broadcasts, $r_{i}(s)$ is the number of neighbors of $i$ that successfully receive its message, and $\abs{N(i)} - r_{i}(s)$ is the number of neighbors of $i$ that are either broadcasting or receiving multiple messages, and thus result in a failure for $i$. With this notation, we can define the reception capacity game. 
\begin{definition}
For constants $c,d \geq 1$, an instance of the \emph{reception capacity game} is given by a graph $G = (V,E)$. The utility for player $i$ is $u_{i}:S \to \bbZ$, defined by $u_{i}(s) = c \cdot r_{i}(s) - d \cdot (\abs{N(i)} - r_{i}(s))$ if $s_{i} = 1$, and $u_{i}(s) = 0$ otherwise. 
\end{definition}
This game intuitively models the fact that each node would like to send its message to its neighbors, and gets a benefit proportional to the number of successes but with a penalty for failures (possibly due to either the cost of wasting the transmission power, or more altruistically, a payment for the interference caused). The parameters $c$ and $d$ provide a means by which to model a difference between the reward of a successful broadcast and the cost of a failure (in the simplest case we can think of $c=d=1$). 


\ifsubmission
\else
\begin{definition}
A \emph{coarse correlated equilibrium (CCE)} is a distribution over $S$ such that in expectation, no player has any incentive to deviate.  Formally, $p$ is a CCE if for any $i \in [n]$ and any $s_{i}' \in \set{0,1}$, 
$\ev{s \sim p}{u_{i}(s)} \geq \ev{s \sim p}{u_{i}(s_{-i}, s_{i}')}$, 
where $s_{-i},s_{i}'$ is a vector formed by replacing the $i$'th coordinate of $s$ with $s_{i}'$.  
\end{definition}

Clearly any Nash equilibrium is a CCE, but a CCE is not necessarily a Nash since every Nash is a product distribution over $S$ while a CCE does not have to be a product distribution.  This definition can be relaxed to that of an approximate CCE. In particular, we say that $p$ is an $\epsilon$-CCE if for any $i \in [n]$ and any $s_{i}' \in \set{0,1}$, it holds that $\ev{s \sim p}{u_{i}(s)} \geq \ev{s \sim p}{u_{i}(s_{-i}, s_{i}')} - \epsilon$. 
Note that a true CCE is a $0$-CCE, and in the reception capacity game, every distribution over $S$ is a $(c+d)n$-CCE.
\fi

\subsection{Notation}
In the following sections, we let $G = (V,E)$ with $V = [n]$ be an instance of the reception capacity game. Without loss of generality we assume that $G$ is connected, since the results below directly extend to the case where $G$ is disconnected by applying the upper and lower bound to each connected component.
We next define a few important sets and quantities. 

For every $s \in S$, we will let $B(s)= \{i \in [n] : s_i = 1\}$ be the vertices which are broadcasting, $R(s)= \{i \in [n] : s_i = 0 \land \sum_{j \in N(i)} s_j = 1\}$ be the vertices which are successfully receiving a transmission, $F(s)= \{i \in [n] : s_i = 0 \land \sum_{j \in N(i)} s_j \geq 2\}$ be the vertices which are receiving at least two transmissions (and thus are failing to successfully receive any transmission), and $A(s) = \{i \in [n] : s_i = 0 \land \sum_{j \in N(i)} s_j = 0\}$ be the vertices which are neither broadcasting nor receiving any message.  

Let $p$ be a distribution over $S$ which is an $\epsilon$-CCE. Note that $\epsilon \geq 0$ without loss of generality, since if $\epsilon \leq 0$ then we are at a true CCE and so are at a $0$-approximate CCE.  With respect to $p$, we can define $B = \sum_{s \in S} p(s) |B(s)|$ as the expected number of broadcasters, $R = \sum_{s \in S} p(s) |R(s)|$ as the expected number of successful receptions (note that this is the quantity which we are trying to compare to \opt), $F = \sum_{s \in S} p(s) |F(s)|$ as the expected number of failures, and $A = \sum_{s \in S} p(s) |A(s)|$ as the expected number of nodes who neither broadcast nor hear a transmission.


\subsection{Lower Bound on Successful Receptions} \label{sec:pota}
%

\ifsubmission
In this section we give a proof sketch of Theorem~\ref{thm:pota} by showing a lower bound on the expected number of successful receptions in any $\epsilon$-CCE, i.e., showing that the quality of any CCE is not too far from \opt. 
All proofs are in Appendix~\ref{app:pota}.
\else
In this section we prove Theorem~\ref{thm:pota} by showing a lower bound on the expected number of successful receptions in any $\epsilon$-CCE, i.e., showing that the quality of any CCE is not too far from \opt. 
\fi

We begin with some lemmas that let us relate $B$ and $F$ to $R$, but for which we need some more notation.  Recall that for every $s \in S$ and $i \in [n]$, we defined $r_i(s) = |R(s) \cap N(i)|$ if $s_i = 1$ and $r_i(s) = 0$ if $s_i = 0$.  In other words, if $i$ is broadcasting in $s$, then $r_i(s)$ is the number of its neighbors that successfully receive its message, and otherwise $r_i(s)$ is $0$.  Similarly, let $f_i(s) = |F(s) \cap N(i)|$ if $s_i = 1$ and let $f_i(s) = 0$ if $s_i = 0$, and let $b_{i}(s) = \abs{B(s) \cap N(i)}$ if $s_{i} = 1$ and let $b_{i}(s) = 0$ if $s_{i} = 0$. 

Since $p$ is an $\epsilon$-CCE we know that every vertex $i$ gets expected utility that is at least $-\epsilon$, since otherwise it would have incentive to get utility $0$ by never broadcasting. 
The expected utility of vertex $i$ under $p$ is precisely $\sum_{s \in S} p(s) \left(c \cdot r_i(s) - d(f_i(s) + b_{i}(s)) \right)$, since if $s_i = 0$ then $c \cdot r_i(s) - d (f_i(s) + b_{i}(s)) = 0$ which is the utility obtained by $i$ by not broadcasting, while if $s_i = 1$ then $r_{i}(s)$ is exactly the number of neighbors that successfully receive $i$'s message, and $f_{i}(s) + b_{i}(s)$ is the number of neighbors of $i$ that are either broadcasting or receiving multiple messages, and thus do not successfully receive $i$'s transmission. Thus, for every $i \in [n]$ it holds that 
\begin{align} \label{eq:CCE-main}
\sum_{s \in S} p(s) \left(c \cdot r_i(s) - d(f_i(s) + b_{i}(s))\right) \geq -\epsilon. 
\end{align}

We proceed by using this to relate $B$ and $F$ to $R$.

\begin{lemma} \label{lem:B}
$B \leq \frac{c+d}{d} \cdot R + \frac{\epsilon n}{d}$. 

\label{lem:pota-b}
\end{lemma}
\ifsubmission
\else
\begin{proof}
For any $s \in S$ and $i \in [n]$, clearly if $s_i = 1$ then $r_i(s) + f_i(s) + b_{i}(s) = |N(i)|$ (since every neighbor of $i$ either successfully receives $i$'s transmission or fails because it is broadcasting or also receiving another transmission). Therefore, \eqref{eq:CCE-main} implies that $\sum_{s \in S} p(s) s_i \left(c \cdot r_i(s) - d(|N(i)| - r_i(s))\right) \geq -\epsilon$, and thus $\sum_{s \in S} p(s) s_i \left((c+d) r_i(s) - d\right) \geq -\epsilon$ (since $|N(i)| \geq 1$).  Rearranging, we get that
\begin{align*}
(c+d)\sum_{s \in S} p(s) r_i(s) = (c+d)\sum_{s \in S} p(s) s_i r_i(s)  \geq d \sum_{s \in S} p(s) s_i - \epsilon. 
\end{align*}
We can now use this to bound the expected number of broadcasters:
\begin{align*}
B &= \sum_{s \in S} p(s) |B(s)| 
= \sum_{s \in S} p(s) \sum_{i=1}^n s_i 
= \sum_{i=1}^n \sum_{s \in S} p(s) s_i \\ 
&\leq \sum_{i=1}^n \left(\frac{c+d}{d} \sum_{s \in S} p(s) r_i(s) + \frac{\epsilon}{d}\right)\\
& = \frac{c+d}{d} \sum_{i=1}^n \sum_{s \in S} p(s) r_i(s) + \frac{\epsilon n}{d} 
= \frac{c+d}{d} \sum_{s \in S} p(s) \sum_{i=1}^n r_i(s) + \frac{\epsilon n}{d}.
\end{align*}
Since every successful reception can be uniquely attributed to a single transmitter, we know that $\sum_{i=1}^n r_i(s) = |R(s)|$ for every $s \in S$.  Thus we get that $B \leq \frac{c+d}{d} \sum_{s \in S} p(s) |R(s)| + \frac{\epsilon n}{d} = \frac{c+d}{d} \cdot R + \frac{\epsilon n}{d}$, as claimed.
\end{proof}
\fi

\begin{lemma} \label{lem:F}
$F \leq \frac{c}{2d} \cdot R + \frac{\epsilon n}{2d}$. 
\label{lem:pota-f}
\end{lemma}
\ifsubmission
\else
\begin{proof}
For any $s \in S$, note that  every failure is due to a collision between at least two messages.  Thus $|F(s)| \leq \frac12 \sum_{i=1}^n f_i(s)$.  Moreover, we know from \eqref{eq:CCE-main} that $\sum_{s \in S}  p(s) \cdot d \cdot f_i(s) \leq \sum_{s \in S} p(s) (c \cdot r_i(s) - d \cdot b_{i}(s)) + \epsilon \leq \sum_{s \in S} p(s) \cdot c \cdot r_i(s) + \epsilon$ for all $i \in [n]$.  Putting this together, we get that
\begin{align*}
F &= \sum_{s \in S} p(s) |F(s)| \leq \frac12 \sum_{s \in S} p(s) \sum_{i=1}^n f_i(s) 
= \frac12 \sum_{i=1}^n \sum_{s \in S} p(s) f_i(s) \\
& \leq \frac12 \sum_{i=1}^n \frac1d \left(\sum_{s \in S} c \cdot p(s) r_i(s) + \epsilon\right)
= \frac{c}{2d} \sum_{i=1}^n \sum_{s \in S} p(s) r_i(s) + \frac{\epsilon n}{2d}\\
&= \frac{c}{2d} \sum_{s \in S} p(s) \sum_{i=1}^n r_i(s) + \frac{\epsilon n}{2d} 
= \frac{c}{2d} \sum_{s \in S} p(s) |R(s)| + \frac{\epsilon n}{2d}
= \frac{c}{2d} R + \frac{\epsilon n}{2d}. \qedhere
\end{align*}
\end{proof}
\fi

The quantity $A$ is more difficult to bound, and will require us to split the nodes into two sets: nodes with large contribution to $A$, and nodes whose contribution to $A$ is bounded. In particular, define $a = (d + \frac{2}{3}c + \epsilon)/(c+d)$, which will be the threshold. Let $X = \set{i \in [n] : \sum_{s \in S : i \in A(s)} p(s) > a}$
be the nodes which contribute a significant amount to $A$. Note that if $\epsilon > \frac13 c$ then $X$ is empty.  Let $Y = V \setminus X$. Let $d_i^X = |N(i) \cap X|$ and let $d_i^Y = |N(i) \cap Y|$. 
We begin with a simple equation which follows directly from the fact that $p$ is an $\epsilon$-CCE. 

\begin{lemma} \label{lem:CCE-eq}
For every $i \in [n]$, it holds that 
$$c \cdot \sum_{j \in N(i)} \sum_{\substack{s \in S: \\ s_{i} = 0 \land j \in A(s)}} p(s) \leq d \cdot \sum_{j \in N(i)} \sum_{\substack{s \in S: \\ s_{i} = 0 \land j \not\in A(s)}} p(s) + \epsilon.$$
\end{lemma}
\ifsubmission
\else
\begin{proof}
Let $i \in [n]$. Note that for any $s \in S$ with $s_i = 0$, if node $i$ were to transmit, then every neighbor in $A(s)$ would result in a successful reception while every neighbor that is not in $A(s)$ (i.e., every neighbor in $B(s) \cup R(s) \cup F(s)$) would result in a failed reception. Formally, we have that $\ev{s \sim p}{u_{i}(s_{-i},1)} = \sum_{s \in S : s_{i} = 0} p(s) (c \cdot \abs{N(i) \cap A(s)} - d \cdot \abs{N(i) \setminus A(s)}) \ + \sum_{s \in S : s_{i} = 1} p(s) u_{i}(s)$, and the second summations is equal to $\ev{s \sim p}{u_{i}(s)}$,
because $i$ gets utility $0$ if $s_i = 0$.  By the definition of an $\epsilon$-CCE we know that $\ev{s \sim p}{u_{i}(s_{-i},1)} - \ev{s \sim p}{u_{i}(s)} \leq \epsilon$, and thus $c \cdot \sum_{s \in S : s_{i} = 0} p(s) \abs{N(i) \cap A(s)} \leq d \cdot \sum_{s \in S : s_{i} = 0} p(s) \abs{N(i) \setminus A(s)}) + \epsilon$. 
Rearranging each sum gives the lemma.  
\end{proof}
\fi

Now we can use this lemma to prove some relationships between $X$ and $Y$.

\begin{lemma}\label{lem:di}
For every $i \in [n]$, it holds that $d \cdot d_{i}^{Y} \geq d_{i}^{X}(a(c+d) - d) - \epsilon$. 
\end{lemma}
\ifsubmission
\else
\begin{proof}
We bound both sides of the inequality in Lemma~\ref{lem:CCE-eq}. First, we have that 
\begin{align*}
\sum_{j \in N(i)} \sum_{\substack{ s \in S: \\ s_i = 0 \\ \land j \in A(s)}} p(s) 
&\geq \sum_{j \in N(i) \cap X} \sum_{\substack{ s \in S: \\ s_i = 0 \\ \land j \in A(s)}} p(s) \\
&= \sum_{j \in N(i) \cap X} \sum_{\substack{s \in S: \\ j \in A(s)}} p(s) 
\geq \sum_{j \in N(i) \cap X} a = a \cdot d_i^X, 
\end{align*}
where we used the fact that for $j \in N(i)$, if $j \in A(s)$ for some $s \in S$ then $s_{i} = 0$. On the other hand, 
\begin{align*}
\sum_{j \in N(i)} \sum_{\substack{s \in S : \\ s_i = 0 \land j \not\in A(s)}} p(s) &= \sum_{j \in N(i) \cap X} \sum_{\substack{s \in S : \\ s_i = 0 \land j \not\in A(s)}} p(s) + \sum_{j \in N(i) \cap Y} \sum_{\substack{s \in S : \\ s_i = 0 \land j \not\in A(s)}} p(s) \\
&\leq \sum_{j \in N(i) \cap X} \sum_{\substack{s \in S : \\ j \not\in A(s)}} p(s) + \sum_{j \in N(i) \cap Y} 1 \\ 
& < \sum_{j \in N(i) \cap X} (1-a) + d_i^Y = (1-a) \cdot d_i^X + d_i^Y.
\end{align*}
Therefore, we can combine these two inequalities with Lemma~\ref{lem:CCE-eq} to get that $ac \cdot d_{i}^{X} \leq d(1-a) d_{i}^{X} + d \cdot d_{i}^{Y} + \epsilon.$
Therefore, we get that $d \cdot d_{i}^{Y} \geq d_{i}^{X}(a(c+d) - d) - \epsilon$, 
which concludes the proof. 
\end{proof}
\fi


\begin{lemma} \label{lem:dix}
Let $i \in X$.  Then $d_i^Y \geq 1$.
\end{lemma}

\ifsubmission
\else
\begin{proof}
Let $i \in X$. Suppose that $d_{i}^{Y} = 0$. Lemma~\ref{lem:di} then implies that $\epsilon \geq \abs{N(i)}(a(c+d) - d) \geq \frac{2}{3}c + \epsilon$ because $\abs{N(i)} \geq 1$, which is a contradiction. 
\end{proof}
\fi

With these lemmas we can now show that $R$ must be large.
\begin{theorem} \label{thm:R}
$R \geq \Omega\left(\frac{cd}{(c+d)^{5/2}} \sqrt{n} - \frac{\epsilon}{c+d} n \right)$.
\end{theorem}
\begin{proof}
The theorem is trivially true when $\epsilon \geq \frac16 c$, since for sufficiently large $n$ the right hand side becomes negative.  Thus we will assume that $\epsilon < \frac16 c$ (which implies that $a < 1$).  

Our first step is to bound $|Y|$. We get that 
\begin{align*}
n &= |Y| + |X| \leq  |Y| + \sum_{i \in X} d_{i}^{Y} = |Y| + \sum_{i \in Y} d_i^X & \text{(by Lemma~\ref{lem:dix})} \\
&\leq  |Y| + \sum_{i \in Y} \left(\frac{d}{a(c+d) - d} \cdot d_i^Y + \frac{\epsilon}{a(c+d) - d} \right) & \text{(by Lemma~\ref{lem:di})} \\
&\leq \frac{a(c+d) + \epsilon}{a(c+d) - d} \abs{Y}^{2} 
= \frac{\frac{2}{3}c + d + 2\epsilon}{\frac{2}{3}c + \epsilon} \abs{Y}^{2} 
< O\left((c + d) \abs{Y}^{2}\right)
\end{align*}
and thus $|Y| \geq \Omega\left(\sqrt{\frac{n}{c+d}}\right)$.  We now relate $|Y|$ to $R$. Note that for every node $i \in Y$, it holds that $\sum_{s \in S: i \not\in A(s)} p(s) > 1-a$. Thus 
\begin{align*}
(1-a)|Y| & \leq \sum_{i \in Y} \sum_{s \in S: i \not\in A(s)} p(s) 
\leq \sum_{i \in [n]} \sum_{s \in S: i \not\in A(s)} p(s) 
= \sum_{s \in S} (n - \abs{A(s)}) \cdot p(s) \\
&= \sum_{s \in S} (|B(s)| + |R(s)| + |F(s)|) \cdot p(s) 
= B + R + F \\
&\leq \left(\frac{c+d}{d} + 1 + \frac{c}{2d}\right)R + \frac{2\epsilon n}{d} 
= \frac{3c + 4d }{2d} \cdot R + \frac{2\epsilon n}{d}, 
\end{align*}
where we use Lemma~\ref{lem:B} to bound $B$ and Lemma~\ref{lem:F} to bound $F$. 
Therefore 
$$R = \Omega\left(\frac{d(1-a)}{c+d} \abs{Y} - \frac{\epsilon n}{c+d} \right) = \Omega\left(\frac{cd}{(c+d)^{2}} \abs{Y} - \frac{\epsilon n}{c+d} \right) = \Omega\left(\frac{cd}{(c+d)^{5/2}} \sqrt{n} - \frac{\epsilon}{c+d} n\right),$$ 
as claimed. 
%
\end{proof}

This immediately gives Theorem~\ref{thm:pota}: since $\opt \leq n$ and $c$ and $d$ are constants, Theorem~\ref{thm:R} implies that $R \geq \Omega\left(n\left(\frac{1}{\sqrt{n}} - \epsilon\right)\right) \geq \Omega\left(\opt\left(\frac{1}{\sqrt{n}} - \epsilon\right)\right)$.

\subsection{Upper Bound on Successful Receptions}
\label{sec:pots}
\ifsubmission
We prove Theorem~\ref{thm:pots} by showing a specific instance of the reception capacity game in which any $\epsilon$-CCE has at most $O((c+d+\epsilon)\sqrt{n})$ expected successful receptions but where the reception capacity is $\Omega(n)$. 
For any $q \in \bbN$, let $G_q = (V_q, E_q)$ be a graph defined as follows.  Let $K = \{v_1, v_2, \dots, v_{3cq+1}\}$, and  for $i \in [3cq+1]$ let $L_i = \{u^i_1, u^i_2, \dots, u^i_{dq}\}$.  Let $L = \cup_{i \in [3cq+1]} L_i$.  We let $V_q = K \cup L$.  Note that the number of nodes $n_q$ is equal to $dq(3cq+1) + 3cq+1 = \Theta(cdq^2)$.  We let the edge set be $E_q = \{ \{v_i, v_j\} : i,j \in [3cq+1], i \neq j\} \cup \{ \{v_i, u^i_j\} : i \in [3cq+1], j \in [dq]\}$.  In other words, we make $K$ a clique, and every vertex in $L_i$ is adjacent only to $v_i$.  The following two lemmas immediately imply Theorem~\ref{thm:pots}. 

\begin{lemma}\label{lem:pots_lb}
The reception capacity of $G_q$ is $\Omega(n_q)$. 
\end{lemma}

\begin{proof}
If every vertex in $K$ broadcasts, then every vertex in $L$ receives a message. Thus the reception capacity is at least $|L| = (3cq+1)(dq) = \Omega(n_q)$.
\end{proof}

\begin{lemma}\label{lem:pots_ub}
In any $\epsilon$-CCE of the reception capacity game in $G_q$, the expected number of successful receptions is at most $O((c+d+\epsilon)\sqrt{n_q}).$ 
\end{lemma}
\begin{proof}
Let $p$ be an arbitrary $\epsilon$-CCE of the reception capacity game in $G_q$.  Then we want to bound $R = \sum_{s \in S} p(s) |R(s)|$.  It is easy to see that a vertex in $L_i$ successfully receives a message if and only if it does not broadcast and vertex $i \in K$ does broadcast, and thus 
$\sum_{s \in S} p(s) |R(s)| \leq \sum_{s \in S} p(s) \left(|K| + \sum_{i \in K} |L_i| s_i\right) 
= 3cq +1 + dq \sum_{i \in K} \sum_{s \in S} p(s) s_i$,
so we just need to bound $\sum_{i \in K} \sum_{s \in S} p(s) s_i$. 

To do this,  we partition the strategy vectors into ``good" vectors (where $i$ might have positive utility), ``bad" vectors (where $i$ has negative utility), and ``irrelevant" vectors (where $i$ has zero utility).  Formally, we partition $S$ into the following three sets: $G_i = \{s \in S: s_i = 1 \land \sum_{j \in K} s_j = 1\}$, $B_i = \{s \in S : s_i = 1 \land \sum_{j \in K} s_j \geq 2\}$, and $I_i = \{s \in S : s_i = 0\}$. If $s \in G_i$ then $i$ broadcasts a message which is successfully heard by all $K$ and by at most all nodes in $L_i$. On the other hand, if $s \in B_i$, then $i$ broadcasts a message which may be heard successfully by all nodes in $L_i$ but which results in a failure at all nodes in $K \setminus \{i\}$.  Thus the expected utility of $i$ under $\epsilon$-CCE $p$ is at most $\sum_{s \in G_i} p(s) (3cq + dq)c + \sum_{s \in B_i} p(s) (- 3cqd + dqc) = (3c+d) cq\sum_{s \in G_i} p(s) - 2cqd\sum_{s \in B_i} p(s)$. 

Since $p$ is an $\epsilon$-CCE we know that this expected utility must be at least $-\epsilon$, since $i$ can receive utility 0 by not broadcasting.  Thus we can rearrange to get $\sum_{s \in B_i} p(s) \leq \frac{3c+d}{2d} \sum_{s \in G_i} p(s) + \frac{\epsilon}{2cdq}$.  We can now use this inequality: 
\begin{align*}
\sum_{i \in K} \sum_{s \in S} p(s) s_i &= \sum_{i \in K} \left(\sum_{s \in G_i} p(s) + \sum_{s \in B_i} p(s) \right) 
\leq \sum_{i \in K} \left(\frac{\epsilon}{2cdq} + \frac{3c+3d}{2d} \sum_{s \in G_i} p(s) \right) \\ 
&= \frac{\epsilon \cdot (3cq+1)}{2cdq} + \frac{3c+3d}{2d} \sum_{i \in K} \sum_{s \in G_{i}} p(s) 
\leq \frac{\epsilon \cdot (3cq+1)}{2cdq} + \frac{3(c+d)}{2d}. 
\end{align*}
The last inequality is because $G_i \cap G_j = \emptyset$ for $i, j \in K$ with $i \neq j$ by the definition of $G_i$ and $G_j$, and thus $\sum_{i \in K} \sum_{s \in G_i} p(s) \leq 1$. 

Therefore, we get that $R \leq 3cq + 1 + dq\left( \frac{\epsilon \cdot (3cq+1)}{2cdq} + \frac{3(c+d)}{2d} \right) = O((c+d + \epsilon)q)$. 
Since $n_q \geq 3cdq^2$ we know that $q \leq \sqrt{n_q}$, and thus this shows that any $\epsilon$-CCE has value at most $O((c+d+\epsilon)\sqrt{n_q})$.  
\end{proof}

\else

%
%

We now prove Theorem~\ref{thm:pots} by upper bounding the expected number of successful receptions in any $\epsilon$-CCE in a specific instance of the reception capacity game. 

\begin{theorem}
There exists an instance of the reception capacity game with $R \leq O\left((c+d+\epsilon)\sqrt{n}\right).$ 
\end{theorem}
\begin{proof}
For any $q \in \bbN$, let $G = (V,E)$ be a graph 
composed of $n = dq(3cq+1) + 3cq+1 = 3cdq^2 + dq + 3cq + q + 1$ vertices, defined as follows. Let $V = K \cup L$ and $L = \bigcup_{i \in [3cq+1]} L_{i}$, where $K$ is a clique on $3cq+1$ vertices, and for each $i \in K$, the set $L_{i}$ is an independent set of size $dq$ such that $v_{i}$ is adjacent to each vertex in $L_{i}$. 



We proceed by bounding the value of any $\epsilon$-CCE.  More formally, if $p$ is a distribution over $S$ which is an $\epsilon$-CCE, we need to bound $R = \sum_{s \in S} p(s) |R(s)|$.  It is easy to see that a vertex in $L_i$ successfully receives a message if and only if it does not broadcast and vertex $i \in K$ does broadcast, and thus 
\begin{align*}
\sum_{s \in S} p(s) |R(s)| &\leq \sum_{s \in S} p(s) \left(|K| + \sum_{i \in K} |L_i| s_i\right) 
= \sum_{s \in S} p(s) \left(3cq+1 + \sum_{i \in K} dq s_i\right) \\
& = 3cq +1 + dq \sum_{i \in K} \sum_{s \in S} p(s) s_i,
\end{align*}
so we just need to bound $\sum_{i \in K} \sum_{s \in S} p(s) s_i$. 
To do this,  we partition the strategy vectors into ``good" vectors (where $i$ might have positive utility), ``bad" vectors (where $i$ has negative utility), and ``irrelevant" vectors (where $i$ has zero utility).  Formally, we partition $S$ into the following three sets:
\begin{align*}
G_i &= \{s \in S: s_i = 1 \land \sum_{j \in K} s_j = 1\} \\
B_i &= \{s \in S : s_i = 1 \land \sum_{j \in K} s_j \geq 2\} \\
I_i &= \{s \in S : s_i = 0\}
\end{align*}
If $s \in G_i$ then $i$ broadcasts a message which is successfully heard by all $K$ and by at most all nodes in $L_i$. On the other hand, if $s \in B_i$, then $i$ broadcasts a message which may be heard successfully by all nodes in $L_i$ but which results in a failure at all nodes in $K \setminus \{i\}$.  Thus the expected utility of $i$ under $\epsilon$-CCE $p$ is at most 
\begin{align*}
&\sum_{s \in G_i} p(s) (3cq + dq)c + \sum_{s \in B_i} p(s) (- 3cqd + dqc) \\ 
&= (3c+d) cq\sum_{s \in G_i} p(s) - 2cqd\sum_{s \in B_i} p(s).
\end{align*}

Since $p$ is an $\epsilon$-CCE we know that this expected utility must be at least $-\epsilon$, since $i$ can receive utility 0 by not broadcasting.  Then we can rearrange to get $\sum_{s \in B_i} p(s) \leq \frac{3c+d}{2d} \sum_{s \in G_i} p(s) + \frac{\epsilon}{2cdq}$.  We can now use this inequality to get our desired bound: 
\begin{align*}
\sum_{i \in K} \sum_{s \in S} p(s) s_i &= \sum_{i \in K} \left(\sum_{s \in G_i} p(s) + \sum_{s \in B_i} p(s) \right) \\ 
&\leq \sum_{i \in K} \left(\frac{\epsilon}{2cdq} + \left(\frac{3c+d}{2d} + 1\right)\sum_{s \in G_i} p(s) \right) \\ 
&= \frac{\epsilon \cdot (3cq+1)}{2cdq} + \frac{3c+3d}{2d} \sum_{i \in K} \sum_{s \in G_{i}} p(s) 
\leq \frac{\epsilon \cdot (3cq+1)}{2cdq} + \frac{3(c+d)}{2d}. 
\end{align*}
The last inequality is because $G_i \cap G_j = \emptyset$ for $i, j \in K$ with $i \neq j$ by the definition of $G_i$ and $G_j$, and thus $\sum_{i \in K} \sum_{s \in G_i} p(s) \leq 1$. 
Therefore, we get that 
\begin{align*}
R &\leq 3cq + 1 + dq\left( \frac{\epsilon \cdot (3cq+1)}{2cdq} + \frac{3(c+d)}{2d} \right) \\ 
&= 3cq+1+\frac{\epsilon (3cq+1)}{2c} + \frac32 q(c+d) 
= \frac{9}{2}cq + 1 + \frac{3}{2}q\epsilon + \frac{\epsilon}{2c} + \frac{3}{2} dq. 
\end{align*}
Since $n \geq 3cdq^2$ we know that $q \leq \sqrt{n}$, and thus this shows that any $\epsilon$-CCE has value at most $O((c+d + \epsilon)\sqrt{n}) = O((c+d+\epsilon)\sqrt{n})$.  
\end{proof}

This immediately implies Theorem~\ref{thm:pots} as a corollary. 
\fi

\section{Open Questions}
We hope that this is only the beginning of analyzing the reception capacity of wireless networks.  Many interesting open questions remain, paralleling the work on unicast capacity.  For example, what if we consider restricted classes of graphs, such as unit-disc graphs, which are typically used to model wireless networks?  Does \pds become easier, and are equilibria in the reception capacity game closer to optimum?  
And what happens if we work in the SINR model rather than the graph model?  For the unicast capacity game, \cite{dinitz10} showed that arbitrary graphs are very easy to analyze but the SINR setting is more complicated.  Can we analyze the Price of Anarchy of the reception capacity game in the SINR model?  


\iflncs
\bibliographystyle{splncs04}
\else
\bibliographystyle{abbrv}
\fi
\bibliography{mybib}

\newpage
\appendix

\section{Approximation Algorithm for MaxPDS} 
\label{sec:approx}
In this section, we give an approximation algorithm for \pds. 
Despite the similarities between \pds and \ucp, we remark that degenerate cases prevent us from presenting it as a black-box reduction to \ucp by invoking the approximation algorithm given by Demaine et al.\ for \ucp. Nevertheless, we observe that the classical decay protocol of Bar-Yehuda et al.\ for transmitting in radio networks~\cite{BGI92} yields a simple approximation algorithm for \pds. We note that the resulting algorithm is also a straightforward adaptation of that of~\cite{Demaine08} for \ucp. 

The decay protocol is given in the classical radio broadcasting setting, where transmissions occur over multiple rounds, and there is a subset $B$ of nodes that have already received the message. The decay protocol, for every node in $B$, is the following: for each round $i$, broadcast to all neighbors; then, with probability $\frac{1}{2}$, continue to the next round and otherwise stop transmitting. In~\cite{BGI92}, they observe that for any node $v$, with constant probability there is a round in which exactly one of $v$'s neighbors will broadcast, and thus $v$ will successfully receive the message (with high probability). This can be modified to a single-round protocol by having each node broadcast with probability $1/2^{i}$ for some $i \in [\log(n)]$. By setting $i$ appropriately, we obtain an $O(\log(n))$ randomized approximation algorithm. 

For completeness, we prove the following theorem. 

\begin{theorem} \label{thm:alg}
There is a polynomial time $O(\log(n))$-approximation algorithm for \pds.  
\end{theorem}
\begin{proof}
Let $G = (V,E)$ be an instance of \pds with $|V| = n$. For any set $S$ of vertices, let $f(S) = |D(S)|$ denote the number of perfectly dominated vertices by $S$. Let \alg be an initially empty set and let \opt denote the optimal set of dominating vertices in the above instance. 

Partition the vertices into $\log(n)$ groups $G_{i}$ such that $v \in G_{i}$ if $2^{i} \leq d(v) < 2^{i+1}$. Then there must exist a group $i^{\star}$ such that $\abs{G_{i^{\star}}} \geq \frac{1}{\log(n)} \cdot n \geq \frac{1}{\log(n)} \cdot f(\opt)$ since $f(\opt) \leq n$. 

Our solution \alg is now constructed by randomly adding each vertex $v$ to $\alg$ independently with probability $\frac{1}{2^{i^{\star}}}$ when $i^{\star} > 0$, and with probability $\frac{1}{2}$ when $i^{\star}=0$. 

Let $S \subset V$ be the vertices that are perfectly dominated by \alg. For any vertex $v \in G_{i^{\star}}$, let $d = d(v) \in [2^{i^{\star}},2^{i^{\star}+1})$. Then, the probability that $v$ is perfectly dominated by \alg is the probability that exactly one of $N(v)$ is in \alg and the remaining vertices in $N(v)$ are not in \alg. Since each vertex is chosen to be in \alg independently, when $i^{\star} > 0$ we have that
\begin{align*}
\p{v \in S} &= \left( d \cdot \frac{1}{2^{i^{\star}}} \right)\left( 1 - \frac{1}{2^{i^{\star}}} \right)^{d-1} \geq \left( 1 - \frac{1}{2^{i^{\star}}} \right)^{2^{i^{\star}+1}-1} \\ 
&\geq \left( 1 - \frac{1}{2^{i^{\star}}} \right)^{2^{i^{\star}+1}} \geq \frac{1}{e^{4}}.
\end{align*}
When $i^{\star} = 0$, then $d = 1$ and 
$\p{v \in S} = \left( d \cdot \frac{1}{2} \right)\left( 1 - \frac{1}{2} \right)^{d-1} = \frac{1}{2} \cdot \left(\frac{1}{2}\right)^{0} = \frac{1}{2}$. 
Therefore, 
\begin{align*}
\e{f(\alg)} &= \sum_{v \in V}\p{v \in S} \geq \sum_{v \in G_{i^{\star}}} \p{v \in S} \geq \min\left\lbrace\frac{1}{e^{4}},\frac{1}{2}\right\rbrace\abs{G_{i^{\star}}} \\ 
&\geq \frac{1}{e^{4}\log(n)}f(\opt). 
\end{align*}
Therefore, $\frac{f(\opt)}{\e{f(\alg)}} = O(\log(n))$ as desired. 

Note that while the above algorithm is randomized, it is straightforward to derandomize in polynomial time using the standard method of conditional expectations. 
\end{proof}

\end{document}